\documentclass[conference]{IEEEtran}
\IEEEoverridecommandlockouts
\usepackage{cite}
\usepackage{amsmath,amssymb,amsfonts}
\usepackage{algorithmic}
\usepackage{graphicx}
\usepackage{textcomp}
\usepackage{xcolor}
\usepackage{multicol}
\usepackage{listings}
\usepackage{booktabs}
\usepackage[numbers]{natbib}
\usepackage{subfig}
\usepackage[ruled,vlined, linesnumbered]{algorithm2e}
\definecolor{DarkGreen}{RGB}{1,50,32}
\lstdefinestyle{myScalastyle}{
  frame=tb,
  language=scala,
  aboveskip=3mm,
  belowskip=3mm,
  showstringspaces=false,
  columns=flexible,
  basicstyle={\small\ttfamily},
  numbers=none,
  numberstyle=\tiny\color{gray},
  keywordstyle=\color{blue},
  commentstyle=\color{DarkGreen},
  stringstyle=\color{purple},
  frame=single,
  breaklines=true,
  breakatwhitespace=true,
  tabsize=3,
}

\def\BibTeX{{\rm B\kern-.05em{\sc i\kern-.025em b}\kern-.08em
    T\kern-.1667em\lower.7ex\hbox{E}\kern-.125emX}}
\begin{document}

\title{Harpocrates: A Statically Typed Privacy Conscious Programming Framework}

\author{\IEEEauthorblockN{Sinan Pehlivanoglu}
\IEEEauthorblockA{
\textit{Indiana University Bloomington}\\
 spehliva@iu.edu}
\and
\IEEEauthorblockN{Malte Schwarzkopf}
\IEEEauthorblockA{
\textit{Brown University}\\
malte@brown.edu}
}

\maketitle

\section{Introduction}
 The internet has become a vital part of our everyday lives. With people relying on web application for getting their groceries, booking trips, scheduling doctor's appointments, paying their taxes and more, a significant amount of sensitive, private user data flows through servers every second. Facebook alone generates over 4PB of user data every day \citep{FBDATA}. 
   
  Recent laws such as GDPR \citep{GDPR} that allow the user to control how their data is used and who it is shared with. In large applications, dynamic privacy logic around user data can get very complex, resulting in a control flow that is difficult to reason about and difficult to maintain. GDPR being very young and only adopted by the EU in 2016, the regulations constantly change and evolve. Data licensing deals like advertisement partnership result in the user data travelling through various different applications. The evolving and distributed nature of privacy inevitably leads to human error and unintentional data leaks in complex applications. 
   
   In this paper,  we introduce Harpocrates, a compiler plugin and a framework pair for Scala that binds the privacy policies to the data during data creation in form of oblivious membranes. Harpocrates eliminates raw data for a policy protected type from the application, ensuring it can only exist in protected form and centralizes the policy checking to the policy declaration site, making the privacy logic easy to maintain and verify. Instead of approaching privacy from an information flow verification perspective, Harpocrates allow the data to flow freely throughout the application, inside the policy membranes but enforces the policies when the data is tried to be accessed, mutated, declassified or passed through the application boundary. The centralization of the policies allow the maintainers to change the enforced logic simply by updating a single function while keeping the rest of the application oblivious to the change. Especially in a setting where the data definition is shared by multiple applications, the publisher can update the policies without requiring the dependent applications to make any changes beyond updating the dependency version. 
    
   Harpocrates' membranes are oblivious. This allows the developer to write familiar Scala code in their existing ecosystem and use their favorite libraries without requiring any changes to the external code. The developer enriches the types annotations in the constructor with special Policy mix-ins which modify the constructor, only allowing the instances to be constructed in a policy protected state and thus eliminating the possibility of a policy protected type existing in the raw, unprotected form. The compiler inserts flexible and context-dependent policy checks when the data leaves the application and needs to be declassified. 
   
   Harpocrates is implemented as a Scala compiler plugin with 15 phases. In order to evaluate the effectiveness, we integrated Harpocrates to the core part of Vizion, a project management and planning software for musicians that is built with the micro-service architecture with over 20,000 lines of code. We evaluated Harpocrates in terms of the number of lines that needed to be added and changed for a successful integration, the compile time overhead as well as runtime overhead. The results are presented in Section 6. 
   
   The rest of the paper is organized as follows: Section 2 summarizes related works in privacy enforcement, Section 3 gives an overview of the developer experience using our framework, Section 4 describes the framework design, Section 5 details the internals of the compiler plugin, Section 6 presents the evaluation of Harpocrates integrated into a real world web application and finally Section 7 concludes the paper and discusses future work.

 \section{Related Work}

 \subsection{What in the monad is this?}
 
 \begin{figure*}
     \centering
    \begin{lstlisting}[style=myScalastyle]
    case object PrivacyError extends Throwable 

    class PrivacyM[A](a: A, check: A => Boolean){
      private val data = a
      
      def map[B](f: A => B) : PrivacyM[B] = 
        if(check(data)){
          new PrivacyM(f(data), (b: B) => {val temp = data; check(temp)})
        }else{
          throw PrivacyError
        }
    }
    \end{lstlisting}
     \caption{A Privacy Monad In Scala. Incorporating this definition into the code base would require manually wrapping every sensitive instance of A and then updating the types everywhere A is used.}
 \end{figure*}
 
 One of the prominent approaches a Scala developer might take when facing the problem of privacy policy enforcement, is to use a monadic data structure. A monad has the ability to wrap a piece of data and maintain full control over any additional computations over it. The Typelevel Cats library \citep{CATS} provide abstractions to easily define monadic structures, and Typelevel Discipline \citep{DISC} allows the programmer to test and verify various properties of these algebraic structures, like associativity. Figure 1 shows an example implementation of a primitive, parameterized privacy monad. While this approach is functional, we note a number of issues with it:
 
 \begin{enumerate}
  \item The developer still needs to manually lift the data into this Privacy Monad. This requirement essentially undermines the argument that human error occurs. Given under this model, the raw data still exists, there is nothing stopping the developer from simply using the raw data without any policy protectionå. 
  \item Type classes in Scala are not oblivious. This means that once the developer wraps the data inside the Privacy Monad at the initialization site, they will have to update their type signatures downstream to align. This can get very problematic in cases where when team of developers maintain a data structure and publish it for others to use. They either would publish it (1) wrapped, or (2) unwrapped. The former would require all the dependent teams to go through significant updates. This can get significantly complicated if the data structure was to go through multiple teams, each adding their own conditions. This would require multiple instances of these privacy monads to be composed, likely through Monad transformers, resulting in type signatures that are hard to reason about. The latter, on the other hand, opens it self up to the issue mentioned above. 
  \item This approach is not powerful enough to reason about potential side-effects. This would require the developer to execute the necessary privacy checks on each computations applied through \emph{map} resulting in performance bottlenecks on even simple operations. 
  
\end{enumerate}
 
\subsection{Static IFC}
 
 Information Flow Control has been a widespread approach to Privacy. For example, JFlow \citep{JFLOW} is a annotation based, static information flow extension of Java. JFlow was one of the first privacy solutions to address practical application development and provided support for complex language features such as mutations, subtyping and exceptions. The decentralized label model introduced by JFlow guarantees privacy constraints in a mutual distrust setting. JFlow relies on precise annotations to function correctly. This means that the developer needs to be aware of the privacy-sensitive locations in the code and understand the privacy requirements of the data and symbols that represent the data very well. This results in a significant risk of human error. In addition, as a result of the fully static nature of these labels, JFlow's isn't able to handle dynamic policies that might depend on an external service. 
 
 Polikarpova et. al introduced LIFTY \citep{LIQUID}, a domain specific language for liquid information flow control. Similar to JFlow, LIFTY requires the developer to annotate the sensitive data sources with declarative security policies. These policies are later statically verify the policies. Unlike JFlow, these annotations are done by types. LIFTY represents information flow control using liquid types that allows for easy verification. LIFTY also suggests a provably correct repair for the policies if they can not be verified.The downside of LIFTY is that it requires the programmers to use their custom monad \emph{TIO} requiring large refactors to existing code bases. 
 
 Storm \citep{STORM} also represents policies in terms of refinement types. Storm policies are predicate functions over a row and a user, providing a very fine grained, non-extendable framework. In addition, Storm specifications require a deep understanding of functions and their potential side effects in order to annotate the functions with the correct policies.

 \subsection{Dynamic IFC}
 
 Biachhawat et. al \citep{GRADUAL} observed that static IFC does not scale well in the scenarios where the security label of a variable is not known at compile time. They introduced a gradually typed language with gradual guarantees. This allows the security labels to be gradually refined at runtime. While this reduces the amount of labels the developer has to place, it does not completely eliminate them. This language, while providing some support for values that are not known until runtime, it still lacks support for concurrent operations that fetch the value from external systems. 
 
 Jacqueline \citep{JACQ} was created specifically for database-backed applications. Jacqueline uses a custom ORM, a custom runtime and a custom web framework. The runtime performs different computations based on the based on target user of the output. Requiring custom end to end frameworks make Jacqueline incompatible with existing ecosystems and requires a notable about refactorization and modification to the existing code bases. 
 
 Resin \citep{Resin} asserts data privacy by propagating policy objects along with the data. Resin wraps any application boundary in filter objects. Once a data with some policy object attached to it passes through a filter object, the privacy checks surrounding the data are run and it is only allowed to pass through if the checks pass. Resin also requires a custom Python runtime and it's policy propagation has a very significant overhead.

 \subsection{Contracts}
 
 Racket's contracts are closest to our work. Chaperones and imposters \citep{CHAPERONE} define membranes around data with conditions whose checking are deferred until invocation. A membrane defined around a list, that dictates what elements can be added to that list, will not execute until the code runs and an element is attempted to be appended to the list. These membranes are procedures that propagate through any computation over the data. Racket contracts are oblivious, allowing the membrane wrapped data to be used just like its raw counterpart. Racket contracts are very coarse-grained, designed to accommodate a large set of needs such as dynamic type checking. In turn, this results in a complex and large developer workload, requiring the developer to define how these membranes propagate for a given data structure and through certain functions. This leaves a big possibility for human error, if the developer does not understand the privacy requirements around the entire code base fully. 
 
 \subsection{Other}
 
 ShillDB \citep{SHILL} approaches privacy as access control with capabilities and pushes the enforcement to the query boundaries. ShillDB provides a query language that determines the view of the returned data based on the querying entity. However it does not concern itself with what happens to the data, once it is read into the application.

 \section{Overview}
 
 Vizion is a web based project management application for musicians. It allows them to manage their writing, recording and production process, book tours as well as market their projects by allowing them to send press releases and review requests to the publications that fit their genre. Privacy laws allow the publications to unsubscribe from any mailing list. 
 
 Once a publication unsubscribes from the mailing list of a band, they should no longer receive communications from them. The information whether a publication unsubscribed or not resides in a database and can not be determined at compile time. Currently this is solved by adding proper conditions with if statements into the \emph{send} function before the email is constructed and sent. This condition will require extra context to run, that is what band is sending the email and to whom. Once this information is present, it can make the necessary call to the database to check if the publication has unsubscribed. 
 
 Harpocrates' power revolves around centralized \emph{Policy Class} definitions. Policy classes implement the \emph{Policy} trait and are required to implement the \emph{check} function that defines the declassification conditions for the given data. Figure 2 shows an example Policy Class used in Vizion. This policy enforces a publication's right to unsubscribe from any Band's mailing list under GDPR. 
 
 \begin{figure*}
 \centering
 \begin{lstlisting}[style=myScalastyle]
class PublicationEmailPolicy[F[_]](email: PublicationEmail,
                                  service: PublicationService[F]) 
                                  extends Policy[PublicationEmail] 
                                  {
     val data = email
     def check()(...): Boolean = {
       ...
       val badBands = service.getBlackListedBands(publicationId)
       !badBands.contains(band)
       ...
    }
}
\end{lstlisting}
\caption{An Harpocrates Policy for Securing Publication Email. The compiler will determine the application boundaries and invoke the check function to determine whether data can escape the application under current context.}
\end{figure*}
 
 The \emph{checkExpanded} function checks whether a given Band is contained in the unsubscribed band list, referred to as \emph{badBands} in the code above. This function requires an instance of a \emph{Band} and a \emph{PublicationId} in order to run. Both of these values may vary between different invocation of the check function. In addition, there may be multiple instance of a Band present at the call site. Therefore the developer is required to annotate the specific variable these values need to be fetched from. Figure 3 presents an example \emph{send} function that utilizes these annotations and sends an email through AmazonSES. These annotations do not necessarily need to be within the lexical scope of the function but can reside anywhere in the fluid scope. 
 
 The Policy Classes at the bare minimum needs to accept the data they are protecting, however just like any other Scala class, the constructor can accept as many arguments as needed. The additional arguments are reserved for any static data that may be needed to execute the check function, that does not change between different invocations of the policy checking, such as any external services that need to be queried. The developer distinguishes the data being protected from the remainder of the arguments by assigning it to the \emph{data} variable. 
 
 Once a policy class is implemented, it is mixed into a constructor with the \emph{enforce} keyword. Figure 4 shows how the \emph{PublicationEmailPolicy} is mixed into the \emph{Publication} definition. Any additional parameters such as the required service, should be specified at this point. Adding the policy into the constructor, overrides the default constructor, removing any possibility of constructing this class in it's raw, unprotected form. Any call to the constructor such as \emph{Publication("name1", "email@email.com", ...)} would automatically result in policy protected data. As a result, when the application receives data from an external source such as an external service or a database, the data is automatically protected when read into this data structure and the developer doesn't need to actively think about policy protection or making sure that the data is protected. 

 Figure 5 shows the complete Policy interface. \emph{ScopedController, combine, and} will be explained in detail in Section 3.
 
 Once the developer goes through these three necessary steps: defines the Policy class, updates the constructor with the necessary policy definitions and annotates the required parameters, the compiler injects the necessary checks. The compiler starts by copying all the methods defined on the enclosed type, such as \emph{PublicationEmail} to the Policy Class. This ensures a structural equality between the enclosed type and the policy membrane, allowing the developer to invoke function on the Policy as if it was the enclosed data itself, without requiring any changes to the existing code, such as mapping of the function into the policy container. The copied functions are checked for any potential side-effects and the body is wrapped inside an invocation of the check function if the compiler can not decide the code is side effect free. There are some special cases such as primitives which are unboxed in Scala, where we provide a Policy constructor for the given type out of the box and skip copying over the methods.The compiler accumulates the annotated parameters as it traverses the code and injects an implicit instance of \emph{PolicyArgs} right before a method invocation on the Policy for the check function to execute properly. The compiler also scans the code for any application boundaries such as call into the database or HTTP servers, and injects an invocation of the check function, declassifying the data and allowing it to leave the application, if and only if it is safe to do so. Last but not least, the compiler injects an implicit conversion from the raw type to a policy protected type at the top level. This is required in order to support functions that are in the standard library or in an external library. Because the don't expect a policy wrapped argument, we lift those classes into a Policy and use the copied over counterpart to the invoked method that can accommodate policy wrapped parameters.

 
 
\begin{figure*}
\centering
\lstset{emph={policyArg},emphstyle=\textbf}
\begin{lstlisting}[style=myScalastyle]
 def sendEmail(@policyArg id: PublicationId, @policyArg band: Band, 
   to: PublicationEmail, title: String, html: String): Unit 
    try {
        val client = ....
        val request = new SendEmailRequest()
          .withDestination(...)
          .withMessage(new Message()
              .withBody(new Body()
                  .withHtml(...)
              .withSubject(new Content()
                  .withCharset("UTF-8").withData(title)))
          .withSource(band.email)
      client.sendEmail(request);
    } catch (...) {
      ...
    }
  }
 }
 
\end{lstlisting}
\caption{A function that sends a press release of a band to a publication using Amazon SES. The only change the developer needs to make for the checks to run is to add \textbf{@policyArg} annotations}
\end{figure*}

\begin{figure*}
\begin{lstlisting}[style=myScalastyle]
  case class Publication(
     id: PublicationId,
     name: PublicationName,
     founded: Date,
     email: PublicationEmail enforce PublicationEmailPolicy(publicationService)
   )
\end{lstlisting}
\caption{An example of how a policy is added to the constructor. Any instance of Publication will have its email field protected by the PublicationEmailPolicy}
\end{figure*}
 
\begin{figure}
\begin{lstlisting}[style=myScalastyle]

trait Policy[K]{
  private[policy] val data : K
  protected def check()(implicit args: PolicyArguments, scope: ScopedController): Boolean
  protected def and(that: Policy[K]) : Policy[K]
  def combine(that: Policy[K]) : Policy[K]
  def unsafeUnwrap(reason: String): K
}
\end{lstlisting}
\caption{The full Policy interface}
\end{figure}

\section{Design}
 
 \subsection{Method Dispatch and Oblivious Policies}
 
 Oblivious properties are obtained by achieving structural equality of the Policy class to the wrapped types. This means that the programmer should be able to call a method defined on the original type, on the policy container, with the same set of arguments. In addition to method dispatch, this has two important implications:
 
 \begin{enumerate}
     \item The arguments and the return type may or may not be policy protected. For example, for a method \emph{m}, defined as \emph{def m(a1: Int, a2: Int) : Int = ...}, the programmer should be able to pass a \emph{Policy[Int]} for a1 \textbf{or} a2. 
     
     \item Similarly a raw, non-policy protected type should be able to accept policy-protected arguments. For example the result of \emph{5 + Policy[Int](5)}, \emph{Policy[Int](5) + 5} and \emph{Policy[Int](5) + Policy[Int](5)} should be equivalent to \emph{Policy[Int](10)}
 \end{enumerate}
 
 Scala's reflection library is not powerful enough to dynamically hijack method calls, therefore Harpocrates implements this by copying the functions defined on the wrapped class to the corresponding Policy classes. When the methods are copied over, their bodies are checked for potential side-effects, and if the compiler can not decide with absolute certainty that the body is side-effect free, it is wrapped in an if statement that invokes the check function. Any fields on case classes are transformed to methods when being copied. Since the primitives in Scala are unboxed, it is difficult to carry out this process, therefore we provide type constructors to construct policy classes on primitives.  When the methods are copied over, any arguments and the return type for which there is policy defined in the application, is automatically wrapped. 
 
 In order to satisfy the two points raised above, we leverage Scala's implicit conversions. For every unique Policy class, we inject an implicit conversion from the core type to the policy protected type with the corresponding check function. This allows the compiler to automatically lift any non-sensitive, raw data into a policy-protected data, creating an equivalence between the raw type and the policy wrapped type, allowing raw data to be passed into the copied over methods that expect policy protected instances.  
 
 Last but not least, the compiler needs to align the types on functions that that live outside of Policy classes. A function \emph{def f(a: A) : B} that performs a computation will now need to accept \emph{Policy[A]} and may need to return \emph{Policy[B]}. The compiler will change the argument types if there is a policy class defined for \emph{A} and will patch the return type based on the type of the expression being returned. 
 
 \subsection{Policy Propagation}
 
  Once a piece of data is wrapped by a Policy, it should not escape until it reaches the application boundaries. This means that any computation over the data should not result in declassification. In other words, the result of a computation over policy-protected data should be protected by the same policy. This imposes two notable challenges: 
  
\begin{enumerate}
    \item The check function of a specific policy may require a full view of the original data and rely on information that may not be present in the result of the method invoked on the original data. For example, we can have a case class \emph{User} that is protected by a \emph{UserPolicy} which requires the \emph{UserId} to run the checks. The computation \emph{user.address} will return a address string that will be protected by the same \emph{UserPolicy}. However, the \emph{UserId} is no longer present in the data being protected.
    
    \item In order to copy the adequate methods to the newly generated Policy over the returned data, the policy class definition needs to statically exist before the wiring can take place. For the same \emph{User} with \emph{UserPolicy} mentioned above, a \emph{Policy[UserAddress]} needs to exist at compile time in order to copy the correct methods defined on \emph{UserAddress} over.
\end{enumerate}
 
 Harpocrates handles the first issue by caching the original view within the lexical scope of the check function prior to constructing the new policy. The original data is bound to a unique symbol and all the references to it are updated to this new symbol. Because case classes are immutable in Scala, it is not possible for the local view to be out of sync with the original data, thus this approach is guaranteed to be sound. 
 
 In order to satisfy the second condition, the compiler conducts a deep traversal of the information flow in the application. Starting with the symbols whose types are a Policy, the compiler tracks the computations on this symbol and in turn the returned data. For each computation, the compiler injects a new Policy Class definition into the application with the correct check function. For example, for the \emph{User} class mentioned above, \emph{user.address} results in a \emph{Policy[UserAddress]} with the check function inherited from the \emph{UserPolicy} being injected into the codebase. We refer to these classes as \textbf{Inorganic Policies}.
 
 \subsection{Policy Inheritance And Composition}
 
 Member access on policy protected case classes present an important question. How should the policies compose if both the original data and the member have policies defined on them? Figure 6 shows an example of such definition. It is possible for the member policy to be less restrictive than the parent policy. In such cases, picking the member policy would result in data leakage, thus requiring the policies to be composed through the \textbf{and} operator. However it is also possible for the policies to be disjoint, in which case composing them through \textbf{and} would result in false negatives, denying access to the entities that should be able to access the data.
 
 \begin{figure}
     \centering
     \begin{lstlisting}[style=myScalastyle]
     case class User enforce UserPolicy(
        ...,
        email: UserEmail enforce UserEmailPolicy,
        ...
      )
     \end{lstlisting}
     \caption{A case class definition where the class itself has an attached policy and a field has a separate policy.}
 \end{figure}
 
It is not possible for the compiler to statically decide which composition is the correct one. Therefore we require the programmer to decide the correct composition and precedence. The \emph{compose} function takes in a \emph{Policy} which the programmer can pattern match on to define different behavior for different parent policies. We also provide traits \emph{DominantPolicy, SubmissivePolicy, AndPolicy, OrPolicy}. Extending these will inherit a \emph{compose} function that respectively: always favors the composee, favors the composer, compose the two policies through \emph{and}, compose the two policies through the \emph{or} operator.
 
 \subsection{Contextual Behaviour}
 
 A Policy may need to behave differently depending on context. One good example of this is declassification based on the requesting entity. In Vizion, a publication's email is policy protected to prevent unauthorized bands from accessing it in order to send press releases. However, when the publication owner goes to their own account, they should be able to view and edit their own email address. In this case it is not only unnecessary to run the checks but may not be possible, because the check function requires a Band object as context in order to evaluate, or when running in local or staging environments, the developer may wish to bypass policy checks on logging statements for debugging purposes. 
 
 In order to give the developer flexibility for such cases, we introduce \emph{Scoped Controllers}. Scoped Controllers are a uniform way to pass implicit arguments into the check functions. It is important to note that Scoped Controllers do not automatically override policy behavior. They simply provide a way for the developer to customize the check function behavior by pattern matching on the contents of the controller, which are determined implicitly at the call site. This behavior is crucial in order to maintain centralized auditing of the policies. If \emph{ScopedControllers} automatically altered the policy behaviour, any changes to the policy restrictions may require changes to the \emph{ScopedController}. Because \emph{ScopedControllers} are resolved implicitly, it may be easy for developers to miss them in oversight when making changes to the policy, leading to human error. Figure 7 shows an example of Scoped Controllers used to allow publications access to their own data in Vizion.

 \begin{figure*}
    \centering
     \begin{lstlisting}[style=myScalastyle]
        ...
        private val httpRoutes: HttpRoutes[F] = HttpRoutes.of[F] {
            case GET -> Root / id / "email" =>
             implicit val entityScope = ScopedController(scope=OwnerScope) 
       
        }
        ....
        
        def check()(implicit args: PolicyArgs, scope:      ScopedController): Boolean = {
            ...
            scope.scope match{
                case Owner => true
                case _ =>  ... //run the check
            }
         }
        
     \end{lstlisting}
     \caption{Example use case of a Scoped Controller. A new scoped defined implicitly at the top level, under the end point, can be matched in the check function body to alter the checking behavior.}
 \end{figure*}
 
 The complete policy class is shown in Figure 9.

 \subsection{Limitations}
 
We consider the implications of asynchrony in Policies. One of the motivations behind this paper has been the fact that, Policies very often require information that can only be determined at runtime, through external services and databases. This means that the check function inevitably may have async behaviour which in turn impacts how the compiler needs to inject these checks. As a first pass, we require any Promises inside the check function to be manually resolved however below we lay down a few possibilities for future work. 
 
 \begin{enumerate}
     \item It is possible to provide extensions to Harpocrates that interop with specific effect types such as \emph{Cats IO} or \emph{Scala Futures}
     
     \item We can extend Harpocrates with an \emph{AsyncPolicy} trait where the check function returns \emph{F[Boolean]}. The tagless final pattern would allow integration with any effect type. The high kinded type \emph{F} is polymorphic and can accept any type that adheres to the context bound of \emph{F}. For example, for \emph{F[\_] : Concurrent}, the developer can pass any higher kinded type for which there is a type class \emph{Concurrent[F]} in scope, such as \emph{Cats IO} or \emph{Scala Future}.
     
     Instead of injecting an if statement, the compiler would need to chain the rest of the computation by \emph{mapping} the result of the check. This would in turn require, either the return type of the function to be patched to adhere to the tagless final pattern or giving the developer meaningful type errors for them to manually fix the return types. 
 \end{enumerate}

 We now raise an issue with Harpocrates' integration into a code base that uses tagless final. Figure 8 shows a User class that contains an asynchronous function that uses the tagless final pattern, updates a User's name in the database, and returns the new name if successful. Now imagine that this User class was wrapped inside a Policy. In Section 4.2, we argued that the computations should not declassify data and any value returned should be Policy protected. For a higher kinded type \emph{F[A]} it is unclear if: 
   
   \begin{enumerate}
       \item The complete return value should be policy protected, \emph{Policy[F[A]]}; 
       \item or if the nested value should be protected, \emph{F[Policy[A]]}.
   \end{enumerate}
   
    \begin{figure*}
     \centering
      \begin{lstlisting}[style=myScalastyle]
        class User(...) {
           ...
           def updateUserName(u: String): F[String]{
            database.query(userNameUpdateQuery).with(u)
           }
           ...
          }
        }
     \end{lstlisting}
     \caption{A user class with an async function \emph{updateUserName}}
 \end{figure*}
   
   This is a difficult question to answer. For a pure F, such as \emph{List}, we need to protect the entire collection as the developer might choose to dictate conditions under which a new value can be appended. Whereas for a concurrent type representing a computation such as \emph{Future}, we may not want to wrap the computation itself. This is an open question that we leave for future work.
 
  \begin{figure*}
     \centering
      \begin{lstlisting}[style=myScalastyle]
   class PublicationEmailPolicy[F[_]](email: PublicationEmail,
                                          service: PublicationService[F]) 
                                          extends Policy[PublicationEmail]
                                          with DominantPolicy{
      val data = email
      def check()(implicit args: PolicyArgs, scope: ScopedController): Boolean = {
        def checkExpanded(band: Band, publicationId: PublicationId): Boolean = {
                val badBands = service.getBlackListedBands(publicationId)
                !badBands.contains(band)
        }
        scope.scope match{
            case Owner => true
            case _ => PolicyArguments(checkExpanded _, args[Band], args[PublicationId])
        }
      }
    }
     \end{lstlisting}
     \caption{Complete Email Policy}
 \end{figure*}

 \section{Implementation}
 
   Harpocrates is written in Scala in approximately 3,000 lines of code and the compiler plugin consists of 15 phases. In addition to the compiler plugin, we made minor changes to the core language parser in order to support the  \emph{enforces} keyword. A \emph{case class} constructed with this keyword simply desugars into a regular \emph{case class} with a custom \emph{apply} function that wraps the necessary fields or the final result in the required Policy. The semantics of the language are not altered. Therefore this could alternatively be implemented as a macro annotation, without requiring any changes to the underlying language. 
   
   One of the prominent features of Harpocrates is that it can support types that are imported from libraries. In order to do this, we rely an TASTy\citep{TASTY}, a new interchange format brought on by Scala 3. The dotty compiler for Scala 3 \citep{DOTTY}, compiles all libraries to the TASTy format, which can easily be traversed and modified, as opposed to the bytecode format in Scala 2 and the overall JVM ecosystem. The dotty compiler also provides the necessary meta-programming tools to parse this format at compile time. Because different processes are used to policy-protect user written classes and those that are imported, after compiling a list of policy protected types we take another pass through the code to mark which types are defined in the current project and which aren't. For the definitions that could not be located, we then search for the corresponding TASTy files and parse them. If the a TASTy file can not be located, the compilation terminates with an error. 
   
   \subsection{Limitations}
   
   In this section we cover a number of implementation limitations that we currently do not support and left for future work. 
   
   \begin{itemize}
       \item We currently do not support import renames or duplicate type names.
       \item We current do not provide support for private fields. Any private field in an enclosed type will be assigned a public getter in the Policy wrapper.
       \item We do not support value types such as \emph{case class A(a: 5)}.
       \item We do not support nested classes inside Policy classes.
       \item We do not provide any Java interop. Policy classes will not work with Java classes or interfaces. 
       \item We do not support variables inside Class definitions. This would cause the cached view of the object inside a \emph{check function} to be out of sync with the real data. Any state inside policy protected classes needs to be final.
   \end{itemize}

 \section{Evaluation}
 
   We integrated Harpocrates into Vizion's publicity module. The publicity module allows artist to send press releases, premiere and review requests as well as connect with PR agents who, if hired, would be given access to certain parts of the artist's project. 
   
   Under GDPR, any publication should be able to unsubscribe from the email lists of a band. This should prevent the publication's email from being leaked to the blocked band. A PR agent on the other hand might or might not be allowed to access the email address depending on the band they are acting on behalf of. The publication staff should be able to view their own emails. 
   
   We evaluate Harpocrates in terms of compile time over-head as well as runtime overhead. For both, we measure the metrics for the base application with no checks, the base application with manual checks inserted at application boundaries by the developer and the application with Harpocrates. All experiments were run on a 2020 Macbook Air with the M1 chip and 8GB of RAM running macOS 11.1. 
   
    \begin{figure*}
    \centering
    \begin{tabular}{ |c|c|c|c|} 
     \hline
     \textbf{} & \textbf{Mean (s)} & \textbf{Median (s)}  & \textbf{Std Dev (s)} \\
     \hline
     \textbf{Base App. No Check} & 72.5 & 72 & 0.48 \\ 
      \textbf{Base App. Manual Check} & 75.8& 76 & 1.72 \\ 
     \textbf{Harpocrates} & 81.4 & 81 & 0.48 \\
     \hline
    \end{tabular}
    \caption{Compile Time Results. Hapocrates display 12.4\% overhead over the base application and 7.8\% over the application with checks inserted manually using if statements }
    \end{figure*}
       
    Figure 10 shows the compile time results. It is important to note that the Scala compiler has incremental compilation. All experiments have been run on clean builds. We observe a 12.4\% overhead over the core application and on average 7.8\% over the application with manual checks. 
    
    \begin{figure*}
    \begin{tabular}{ |c|c|c|c|} 
     \hline
     \textbf{} & \textbf{Median Response Time (ms)} & \textbf{Mean (ms)}  & \textbf{99 pct (ms)}  \\
     \hline
     \textbf{Base App. No Check} & 144 & 175 & 196 \\ 
     \textbf{Base App. Manual Check} & 212 & 225 & 237 \\ 
     \textbf{Harpocrates} & 217 & 243 & 255 \\
     \textbf{Base App. Manual Check Ext.} & 1797 & 1925 & 2086 \\ 
     \textbf{Harpocrates Ext.} & 1940 & 2137 & 2273\\
     \hline
    \end{tabular}
    \caption{Run Time Results. Harpocrates display 39\% overhead over the base application and up to 11\% overhead over the application with manual checks using explicit if statements}
 \end{figure*}

   Figure 11 shows the runtime results. We use Gatling \citep{Gatling} to measure run time performance. Each run of the experiment is preceded by 1000 requests in order to normalize the JVM conditions. The warm up is preceded by another 1000 requests and we record the median, mean and the 99th percentile of the response times. We observe a 39\% runtime overhead over the base application with no checks. However, this is expected as each invocation of the check function, essentially requires another call to the database. These calls are inevitable, and necessary whether managed and executed by Harpocrates, or manually by the developer. The downside of the latter is that it requires the developer to carefully think through where the checks needs to be inserted, and thus is prone to human error. The runtime overhead over the manual checks on average is 8\%. This difference occurs because of the extra layer of indirection Harpocractes adds and the more conservative protection it adapts. For example, Harporactes may invoke the checks for logging and/or for calls to the external libraries if it can not decide with certainty that these calls do not leak data outside of the application boundaries. A developer, when manually adding the checks, would likely choose to ignore these. While Harpocrates is conservative by default, it still the provides tools to the developer for more fine-grained optimizations. For example, \emph{Scoped Controller} can allow the developer to bypass checks on all logging by creating their own class around the logging library of their choice and using a scope inside that allows them to bypass the checks. 
   
        \begin{figure}
         \centering
    \begin{tabular}{ |c|c|} 
    
     \hline
     \textbf{} & \textbf{Lines of Code} \\
     \hline
     \textbf{Initial Number of Lines} & 3598 \\ 
     \textbf{Added Lines} & 45 \\ 
     \textbf{Changed Lines} & 17  \\
     \textbf{Total Changes} & 62  \\ 
     \textbf{Final LoC} & 3660 \\
     \hline
    \end{tabular}
    \caption{Initial number of lines of code in Vizion, the number of lines changed and the number of lines added to the application in order to integrate Harpocrates }
 \end{figure}
   
   In our experiments, the application has two notable boundaries for the \emph{send email} endpoint workflow, where the check function is invoked. In order to force more invocations, we introduced eight artificial application boundaries to the publication module. We observe up to 11\% runtime overhead in this conservative setting. 
   
   In order to integrate Harpocrates into the existing Vizion codebase, we had to change 17 lines of code and add 45 new lines, defining the publication email policy. Figure 12 presentes the lines of code before and after Harpocrates. 
    
 \section{Conclusion}
 
 In this paper, we have introduced Harpocrates, a privacy-aware programming framework in Scala that readily integrates with the existing ecosystem of the language. To best of our knowledge, Harpocrates is the first framework with dynamic policy checking that targes a complex, real world language with reasonable compile time and run time overhead. We integrated Harpocrates into a real world application, Vizion and observed an average of 8\% compile time overhead and up to 11\% runtime overhead over a version application with manual checks, in exchange for centralized policy enforcement that is easy to maintain and reason about. 
 
 For future work, we hope to investigate the serializability of Harpocrates policies and extend our work to a distributed setting. In addition, we see the previously raised question of policies in effect systems, and understanding what the policy of a Promise would be, worth exploring.

\bibliography{harp}

\begin{thebibliography}{10}

\bibitem{GRADUAL}
A.~Bichhawat, M.~McCall, and L.~Jia.
\newblock Gradual security types and gradual guarantees.
\newblock In {\em 2021 IEEE 34th Computer Security Foundations Symposium (CSF)}, pages 1--16, 2021.

\bibitem{CATS}
T.~Cats.
\newblock https://typelevel.org/cats/.

\bibitem{DOTTY}
S.~D. Compiler.
\newblock https://github.com/lampepfl/dotty.

\bibitem{DISC}
T.~Discipline.
\newblock https://github.com/typelevel/discipline.

\bibitem{Gatling}
Gatling.
\newblock https://gatling.io/.

\bibitem{STORM}
N.~Lehmann, R.~Kunkel, J.~Brown, J.~Yang, N.~Vazou, N.~Polikarpova, D.~Stefan, and R.~Jhala.
\newblock {STORM}: Refinement types for secure web applications.
\newblock In {\em 15th {USENIX} Symposium on Operating Systems Design and Implementation ({OSDI} 21)}, pages 441--459. {USENIX} Association, July 2021.

\bibitem{JFLOW}
A.~C. Myers.
\newblock Jflow: Practical mostly-static information flow control.
\newblock In {\em Proceedings of the 26th ACM SIGPLAN-SIGACT Symposium on Principles of Programming Languages}, POPL '99, page 228–241, New York, NY, USA, 1999. Association for Computing Machinery.

\bibitem{FBDATA}
J.~W. Nathan~Bronson.
\newblock Facebook’s top open data problems, 2014.

\bibitem{TASTY}
A.~O. of~TASTY.
\newblock https://docs.scala-lang.org/scala3/guides/tasty-overview.html.

\bibitem{LIQUID}
N.~Polikarpova, D.~Stefan, J.~Yang, S.~Itzhaky, T.~Hance, and A.~Solar-Lezama.
\newblock Liquid information flow control.
\newblock {\em Proc. ACM Program. Lang.}, 4(ICFP), aug 2020.

\bibitem{GDPR}
G.~D.~P. Regulation.
\newblock https://gdpr-info.eu/.

\bibitem{CHAPERONE}
T.~S. Strickland, S.~Tobin-Hochstadt, R.~B. Findler, and M.~Flatt.
\newblock Chaperones and impersonators: Run-time support for reasonable interposition.
\newblock In {\em Proceedings of the ACM International Conference on Object Oriented Programming Systems Languages and Applications}, OOPSLA '12, page 943–962, New York, NY, USA, 2012. Association for Computing Machinery.

\bibitem{JACQ}
J.~Yang, T.~Hance, T.~H. Austin, A.~Solar-Lezama, C.~Flanagan, and S.~Chong.
\newblock Precise, dynamic information flow for database-backed applications.
\newblock {\em SIGPLAN Not.}, 51(6):631–647, jun 2016.

\bibitem{Resin}
A.~Yip, X.~Wang, N.~Zeldovich, and M.~F. Kaashoek.
\newblock Improving application security with data flow assertions.
\newblock In {\em Proceedings of the ACM SIGOPS 22nd Symposium on Operating Systems Principles}, SOSP '09, page 291–304, New York, NY, USA, 2009. Association for Computing Machinery.

\bibitem{SHILL}
E.~Zigmond, S.~Chong, C.~Dimoulas, and S.~Moore.
\newblock Fine-grained, language-based access control for database-backed applications.
\newblock {\em Art Sci. Eng. Program.}, 4:3, 2020.

\end{thebibliography}
\bibliographystyle{abbrv}

\end{document}